\newcommand{\g}{$\gamma$}
\newcommand{\cd}{C$_6$D$_6$}

\documentclass[twocolumn]{webofc}
\usepackage{epsfig}
\usepackage{graphicx}
\DeclareGraphicsExtensions{.png,.jpg}
\usepackage[varg]{txfonts}   
\usepackage{booktabs}
\usepackage{array} 
\newcolumntype{L}[1]{>{\raggedright\let\newline\\\arraybackslash\hspace{0pt}}m{#1}}
\newcolumntype{C}[1]{>{\centering\let\newline\\\arraybackslash\hspace{0pt}}m{#1}}
\newcolumntype{R}[1]{>{\raggedleft\let\newline\\\arraybackslash\hspace{0pt}}m{#1}}

\graphicspath{{graphics/}{graphics/arch/}{Graphics/}{./}} 
\begin{document}
\title{New perspectives for neutron capture measurements in the upgraded CERN-n\_TOF Facility}
%
%
\author{%
J.~Lerendegui-Marco\inst{1} \and %
A.~Casanovas\inst{2} \and %
V.~Alcayne\inst{3} \and %
O.~Aberle\inst{4} \and %
S.~Altieri\inst{5} \and %
S.~Amaducci\inst{6} \and %
J.~Andrzejewski\inst{7} \and %
V.~Babiano-Suarez\inst{1} \and %
M.~Bacak\inst{4} \and %
J.~Balibrea\inst{1}
% ...
 and the n\_TOF Collaboration}
\institute{%
Instituto de F\'{\i}sica Corpuscular, CSIC - Universidad de Valencia, Spain \and
Universitat Polit\`{e}cnica de Catalunya, Spain \and
Centro de Investigaciones Energ\'{e}ticas Medioambientales y Tecnol\'{o}gicas (CIEMAT), Spain \and
European Organization for Nuclear Research (CERN), Switzerland \and
Laboratori Nazionali di Pavia, Italy \and
INFN Laboratori Nazionali del Sud, Catania, Italy \and
University of Lodz, Poland
% ...
}

\abstract{%

The n\_TOF facility has just undergone in 2021 a major upgrade with the installation of its third generation spallation target that has been designed to optimize the performance of the two n\_TOF time-of-flight lines. This contribution describes the key features and limitations for capture measurements in the two beam lines prior to the target upgrade and presents first results of (n,\g) measurements carried out as part of the commissioning of the upgraded facility. In particular, the energy resolution, a key factor for both increasing the signal-to-background ratio and obtaining accurate resonance parameters, has been clearly improved for the 20~m long vertical beam-line with the new target design while keeping the remarkably high resolution of the long beamline n\_TOF-EAR1. The improvements in the n\_TOF neutron beam-lines need to be accompanied by improvements in the instrumentation. A review is given on recent detector R\&D projects aimed at tackling the existing challenges and further improving the capabilities of this facility.
 }
\maketitle

\section*{Introduction}\label{sec:intro}
Neutron-induced reactions in general, and more particularly neutron capture (n,\g) reactions, are important in various research fields. Neutron capture reactions play a key role in the design of innovative nuclear devices aimed at the transmutation of nuclear waste, such as accelerator-driven systems, future Generation IV reactor systems. The viability and operation of such systems, featuring fast neutron spectra or new fuel compositions, require an improved knowledge of neutron cross sections~\cite{Salvatores:08,Colonna:10}. As a consequence, more accurate measurements are required, as it is assessed by the WPEC-26 Group of the NEA and its High Priority Request List (HPRL~\cite{Dupont:20}).

Neutron capture reactions play also a fundamental role in the slow neutron capture (s-) process of nucleosynthesis operating in red-giant and massive stars~\cite{Kappeler11,Pignatari:10}, which is responsible for the formation of about half of the elements heavier than iron. The most crucial and difficult data to obtain are the stellar neutron capture (n,\g) cross sections of unstable isotopes which act as branchings of the s-process and yield a local isotopic pattern which is very sensitive to the physical conditions of the stellar environment~\cite{Kappeler11}. At present, there are still about 20 relevant s-process branching point isotopes whose cross section could not be measured yet over the neutron energy range of interest for astrophysics due to limitations of the neutron beam facilities, detection systems and attainable sample masses~\cite{Guerrero:17,Domingo:22}.

The neutron energy range of interest for the measurement of neutron capture cross sections varies depending on the application and hence, pulsed white neutron beams combined with the time-of-flight (TOF) technique, such as the CERN neutron time-of-flight facility (n\_TOF), are the best suited facilities for comprehensive measurements of neutron capture cross sections. 
This work reviews the upgraded capabilities of the CERN n\_TOF facility for (n,\g) measurements after the installation of its third generation spallation target. Sec.~\ref{sec:ntof} discusses the strengths and limitations of the n\_TOF facility for neutron capture measurements before the target replacement. The first experimental results of capture measurements after the facility upgrade are presented in Sec.~\ref{sec:upgraded_target}. Last, Sec.~\ref{sec:upgraded_detectors} describes some recent detector developments that try to cope with experimental challenges and aim at enhancing the detection sensitivity in neutron-capture experiments.

\section{Neutron capture measurements at the CERN n\_TOF facility }\label{sec:ntof}
The n\_TOF facility at CERN generates its neutron beams through spallation reactions of 20~GeV/c protons extracted in pulses from the CERN Proton Synchrotron and impinging onto a lead spallation target. The resulting high energy (MeV-GeV) spallation neutrons are partially moderated in a surrounding water layer to produce a white-spectrum neutron beam that expands in energy from thermal to a few GeV. The neutrons travel along two beam lines towards two experimental areas: EAR1 at 185~m (horizontal)~\cite{Guerrero13} and EAR2 at 19~m (vertical)~\cite{Weiss15}.  A new experimental area, so-called NEAR Station, has been installed during the recent upgrade next to the spallation target. It features an extremely high neutron flux that opens the door to (n,\g) activation measurements on short-lived radioactive isotopes or on very small mass samples~\cite{StamatiND:22,Gervino:22}. 

In neutron TOF capture measurements at n\_TOF the sample of the isotope of interest is placed in the pulsed neutron beam and the prompt capture $\gamma$-rays originating from the sample are registered by means of radiation detectors. Two different detection systems have been used for (n,\g) measurements at n\_TOF so far: the 4$\pi$ n\_TOF Total Absorption Calorimeter (TAC)~\cite{Guerrero:09}--- a segmented detector array consisting of 40 BaF$_{2}$ crystals--- and low-efficiency \cd~liquid scintillators~\cite{Plag03} in conjunction with the pulse-height weighting technique (PHWT)~\cite{Tain02,Tain04}, which allows one to virtually mimic an ideal total energy detector (TED)~\cite{Moxon63}. The latter are more extensively used due to the their reduced neutron sensitivity and fast response.

Since 2001, more than 60 neutron capture cross section measurements have been carried out at the first experimental area n\_TOF-EAR1~\cite{Gunsing:16}. Thanks to its long flight path of 185~m, this beam-line always featured an excellent time-of-flight (i.e. neutron energy) resolution, hence allowing to extend the Resolved Resonance Region (RRR) and extract accurate resonance parameters significantly beyond previous measurements~\cite{Domingo:06_b,Massimi:12,Lerendegui:18,Lederer:19,BabianoND:22}. As a consequence of the long flight path and the the fast recovery of \cd~ detectors from the so-called $\gamma$-flash (i.e. prompt \g-rays produced in the spallation reactions and relativistic particles), EAR1 has provided (n,\g) data in the Unresolved Resonance Region (URR) up to 1~MeV~\cite{Aerts:06,Lederer:11,Mingrone:17}. 

In 2014, the n\_TOF Collaboration built a new vertical beam line, so-called n\_TOF-EAR2~\cite{Weiss15}, with a flight path of only 20 m. The shorter vertical beam line of EAR2 provides a 400 times higher instantaneous neutron flux (see Ref.~\cite{Lerendegui16}), which makes it specially well suited for measuring highly radioactive and/or small mass samples~\cite{Barbagallo:16,Alcayne:20}. 

\begin{figure}[!htb]
  \centering
  \includegraphics[width=\columnwidth]{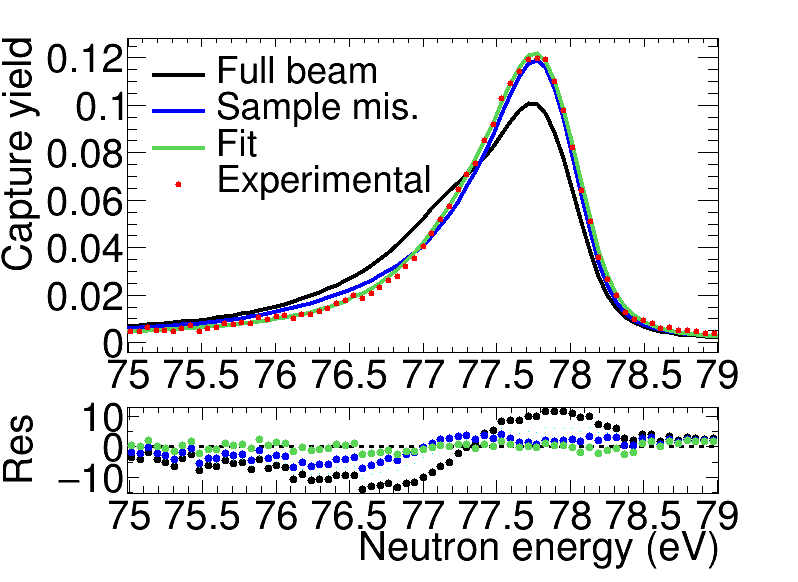}
  \caption{Resonance in the $^{197}$Au(n,\g) yield compared to the initial prediction from MC simulations using the \textit{Full Beam} and an improved fit obtained after considering the actual diameter (5~mm) and miss-alignment (2~mm) of the sample. The curve labeled as \textit{Fit} is based on a parameterization of the energy resolution.}
 \label{fig:RFEAR2_Limitation}
\end{figure}

For example, the recent $^{244,246,248}$Cm(n,$\gamma$) (600~$\mu$g, 1.8~GBq) measurement~\cite{Alcayne:20,AlcayneND:22} has shown the potential of the facility to measure (n,\g) cross sections on radioactive isotopes with low masses. On the other hand, the analysis of this measurement highlighted the difficulties to model the experimental energy broadening  with Monte Carlo (MC) simulations due to the large sensitivity of the resolution to the sample size and position (see Fig.~\ref{fig:RFEAR2_Limitation}). The $^{244,246,248}$Cm(n,$\gamma$) experiment also showed the limited performance of the existing \cd~detectors in the high count-rates of EAR2, as it is discussed in Sec.~\ref{sec:upgraded_detectors}. Indeed, the limited energy resolution of EAR2 and the uncertain experimental conditions motivated that several (n,\g) measurements on unstable nuclei were still carried out in the well-established EAR1~\cite{Guerrero:2020,Casanovas:20}.

Following the main strengths and limitations of the TOF beam-lines described in this section, the recent target upgrade aimed at preserving the high-energy resolution and neutron intensity of EAR1, while significantly optimizing the performance of EAR2, especially in terms of energy resolution.

\section{Improved capabilities after the n\_TOF target upgrade }\label{sec:upgraded_target}
The n\_TOF facility installed its third generation spallation target~\cite{Exposito:21} during the CERN LS2 (2019-2021) with the aim of optimizing the features of both experimental areas, unlike the previous one specifically designed for EAR1. As part of the commissioning of the new target, (n,\g) measurements were carried out mainly aimed at evaluating the energy resolution of the upgraded facility and validating its modelling with MC simulations.  

\begin{figure}[!t]
  \centering
  \includegraphics[width=0.96\columnwidth]{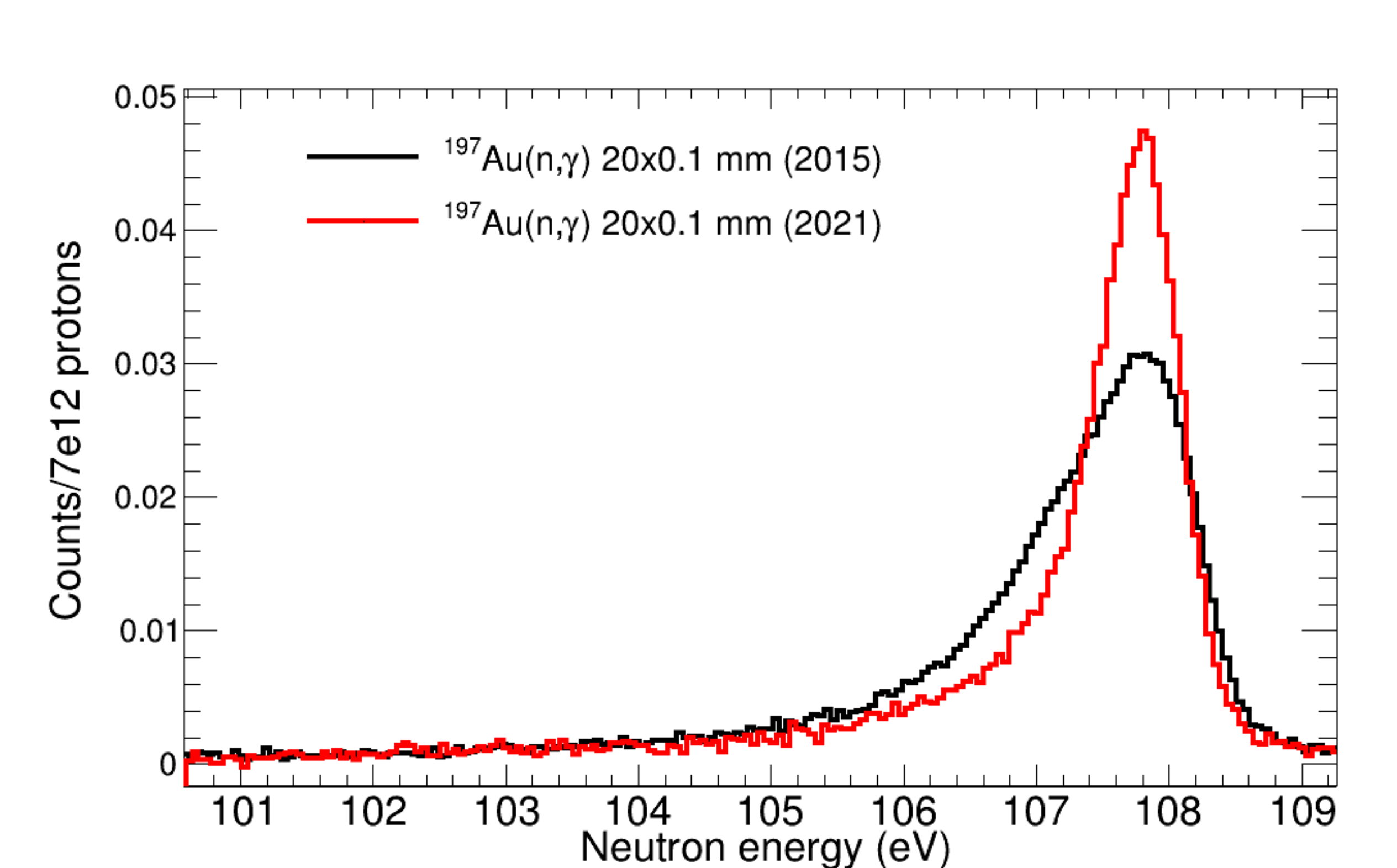}
  \includegraphics[width=1.01\columnwidth]{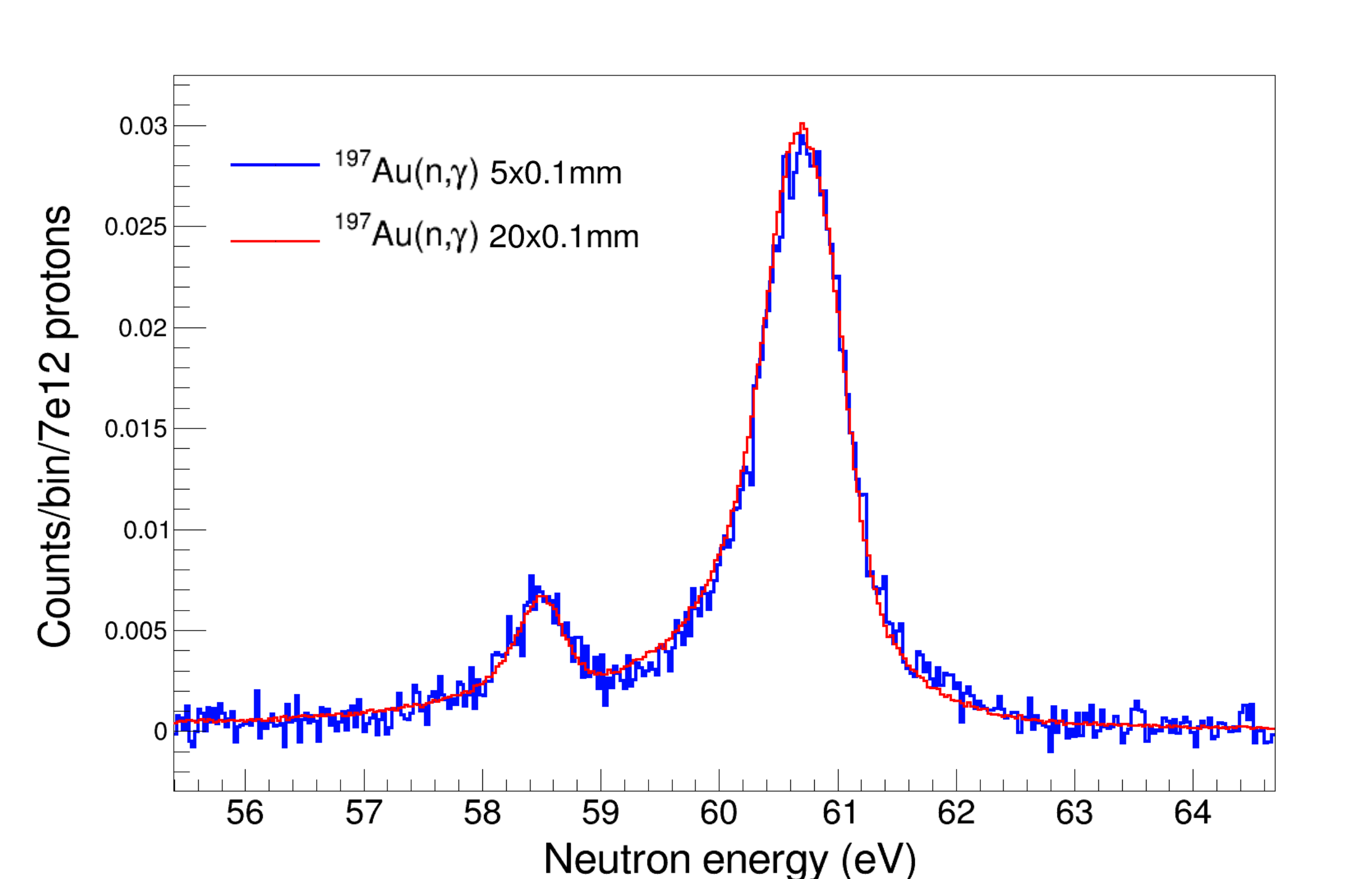}
  \caption{Top: Resonance of the $^{197}$Au(n,\g) yield measured with \cd~detectors at n\_TOF-EAR2
 with the previous (2015) and the upgraded (2021) spallation target. Bottom: Comparison of the yield measured with the new target and two samples of 20 and 5~mm diameter.}
  \label{fig:ImprovedRFEAR2}
\end{figure}

The results indicate that the excellent energy resolution of EAR1 remains unaltered~\cite{BacakND:22}, as expected from the design phase~\cite{Exposito:21}. As for the second experimental area EAR2, the optimization of the spallation target has led to a remarkable reduction in the resonance broadening over the whole studied range from 1~eV to 50 keV (see Fig.~\ref{fig:ImprovedRFEAR2}). Moreover, the measurement of $^{197}$Au samples of various dimensions, shown in Fig.~\ref{fig:ImprovedRFEAR2}, indicates that energy resolution is not sensitive to the sample dimension in contrast to what was found with the previous target (see Fig.~\ref{fig:RFEAR2_Limitation}).  Besides the improved resolution, the upgraded target has also optimized the neutron flux in both experimental areas. According to MC simulations for the new target, the neutron flux in the energy range of interest for (n,\g) measurements (e.g. 10 meV to 1 MeV) is expected to increase in 30-50\% in both EARs. Further details on the characterisation of the neutron flux and preliminary results can be found in Refs.~\cite{BacakND:22,PavonND:22}. 

Optimizing the characteristics of the facility is not sufficient to overcome all the existing challenges for (n,\g) measurements. Indeed, the increase in flux does not help to improve the signal-to-background ratio (SBR) in measurements where the neutron-induced background dominates. In addition, in order to fully exploit the even higher instantaneous flux of EAR2, one requires also of radiation detectors with a fast time-response and a high count-rate capability. New detection systems that cope with these challenges are discussed in Sec.~\ref{sec:upgraded_detectors}.

\section{Experimental limitations and new detection systems }\label{sec:upgraded_detectors}
\subsection{i-TED: suppressing n-induced background via \g-ray imaging}
Despite their low neutron sensitivity~\cite{Mastinu:13}, state-of-the-art \cd~detectors present limited background rejection capabilities. In particular, in many TOF capture experiments a large background component~\cite{Zugec14} arises from scattered neutrons that get subsequently captured in the surroundings of the \cd~detectors. This background has represented the dominant contribution in many (n,\g) experiments in the energy interval of relevance for astrophysics (see e.g. Refs.~\cite{Domingo06,Tagliente13}).

\begin{figure}[!b]
  \centering
  \includegraphics[width=0.9\columnwidth]{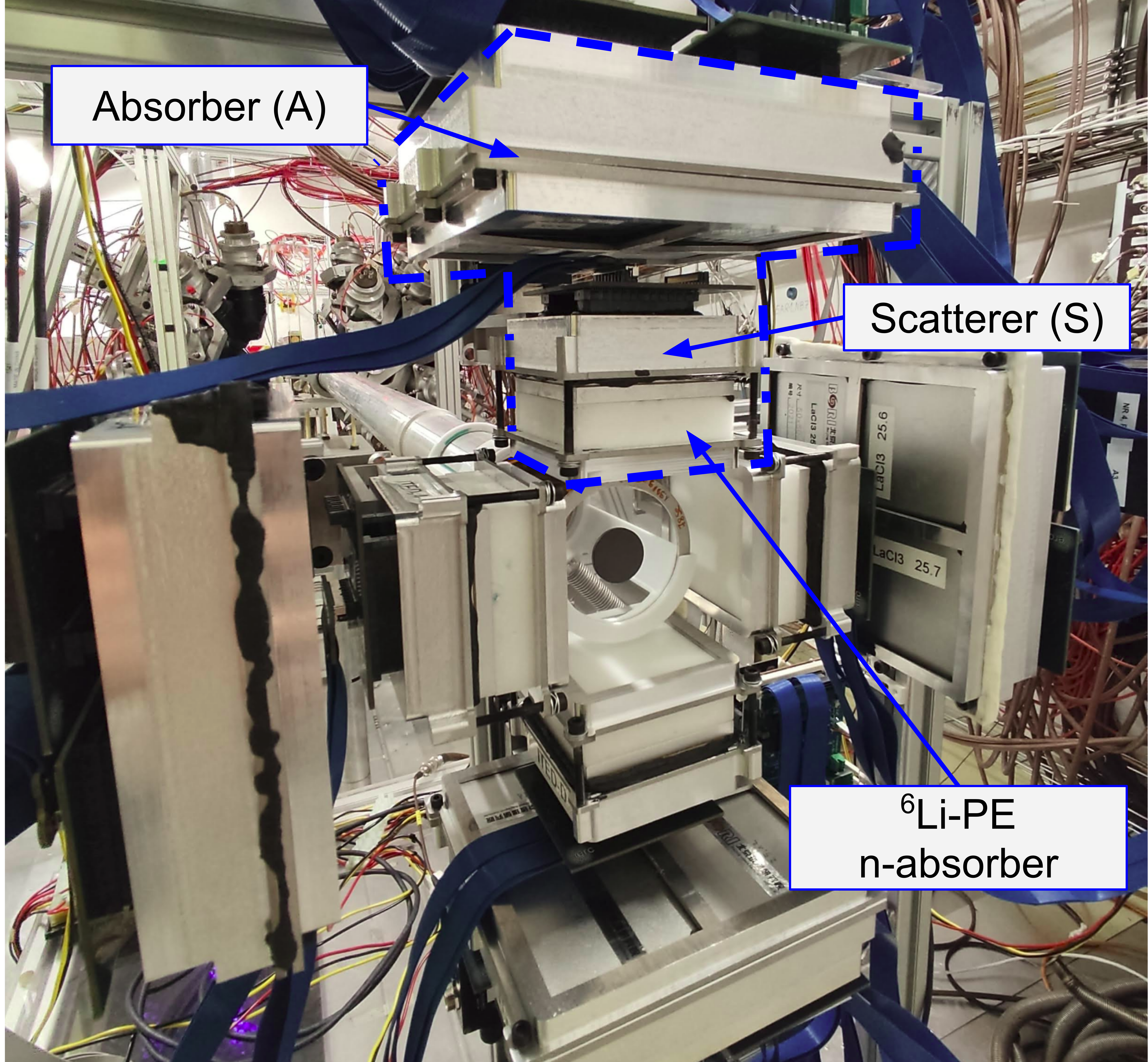}
  \includegraphics[width=1.05\columnwidth]{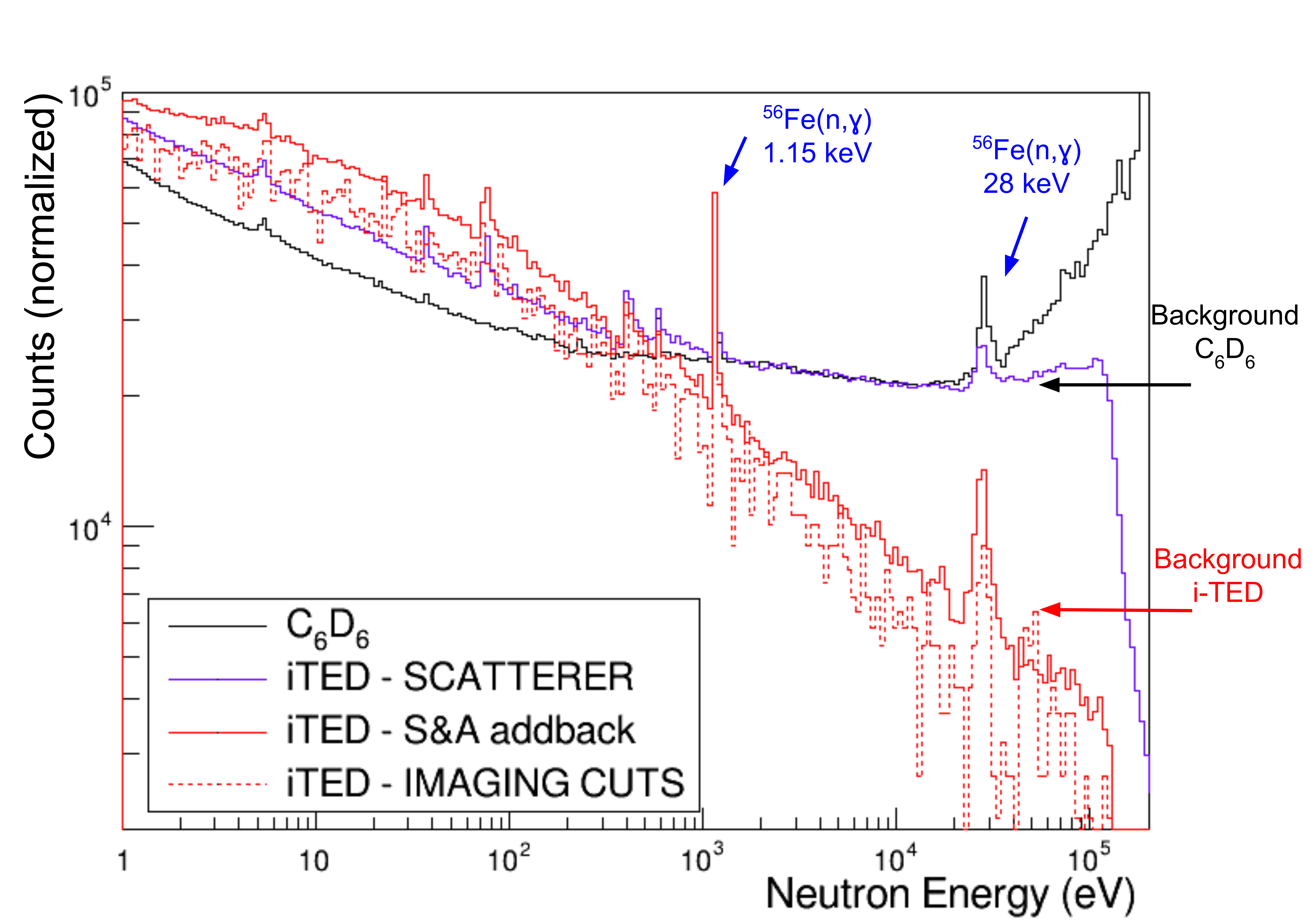}
  \caption{Top: Final i-TED array used in the $^{79}$Se($n,\gamma$) experiment in 2022 at CERN n\_TOF EAR1. One of the Compton modules has been highliged to indicate its basic components. Bottom: $^{56}$Fe(n,$\gamma$) counting rate as function of the neutron energy measured with \cd~detectors and i-TED in the POC experiment of the latter (see text for details).}
  \label{fig:iTED}
\end{figure}

In order to reduce this dominant source of background a total energy detector with $\gamma$-ray imaging capability, so-called i-TED, has been recently proposed~\cite{Domingo16}. i-TED exploits Compton imaging techniques with the aim of determining the direction of the incoming $\gamma$-rays. This allows rejecting events which do not originate in the sample, thereby enhancing the SBR. This novel detection system has been fully developed and optimized in the recent years~\cite{Babiano20,Balibrea:21}. The final i-TED array, shown in the top panel of Fig.~\ref{fig:iTED}, has been used in 2022 at n\_TOF EAR1 for the first measurement of the $^{79}$Se($n,\gamma$) capture cross section~\cite{Lerendegui21}.

The first experimental proof-of-concept (POC) experiment of the background reduction with i-TED was carried out by measuring the $^{56}$Fe(n,$\gamma$) reaction at CERN n\_TOF-EAR1~\cite{Babiano21}. The bottom panel of Fig.~\ref{fig:iTED} shows the measured counts as a function of the neutron energy obtained with \cd~detectors and i-TED. The spectra are normalized to the isolated resonance at a neutron energy of 1.15~keV with the aim of studying the Signal-to-Background ratio (SBR). The results indicate that the SBR at 10~keV is enhanced in a factor 3 when the A- and S-planes are operated in time coincidence, shown in Fig.~\ref{fig:iTED} a solid red line. A maximum SBR of 3.5 (red dashed line) was achieved by applying cuts in the imaging domain~\cite{Babiano21}. These measurements were carried out with a small prototype and, therefore, they were mainly intended to technically validate the system and demonstrate the feasibility of the proposed method. A more quantitative assessment of the i-TED performance will be reported for the full array of 4 Compton modules, as discussed in Ref.~\cite{Babiano21}.

\subsection{s-TED: segmented detection volumes for n\_TOF-EAR2}
The high instantaneous flux of n\_TOF-EAR2, inducing counting rates beyond 10~MHz in the existing \cd~detectors (0.6-1 L volume)~\cite{Plag03,Mastinu:13}, and the intense \g-flash at this facility, led to severe experimental difficulties~\cite{AlcayneND:22}. Among others, counting-rate and time-of-flight dependent variations of the photo-multiplier (PMT) gain and large pile-up effects have been observed. The PMT-related effects have been strongly mitigated with a better choice of PMT~\cite{PMT}. On the other hand, the large volumes of the existing detectors forces one to place them further away from the beam-line to prevent severe pile-up corrections.

\begin{figure}[!h]
  \centering
  \includegraphics[width=0.9\columnwidth]{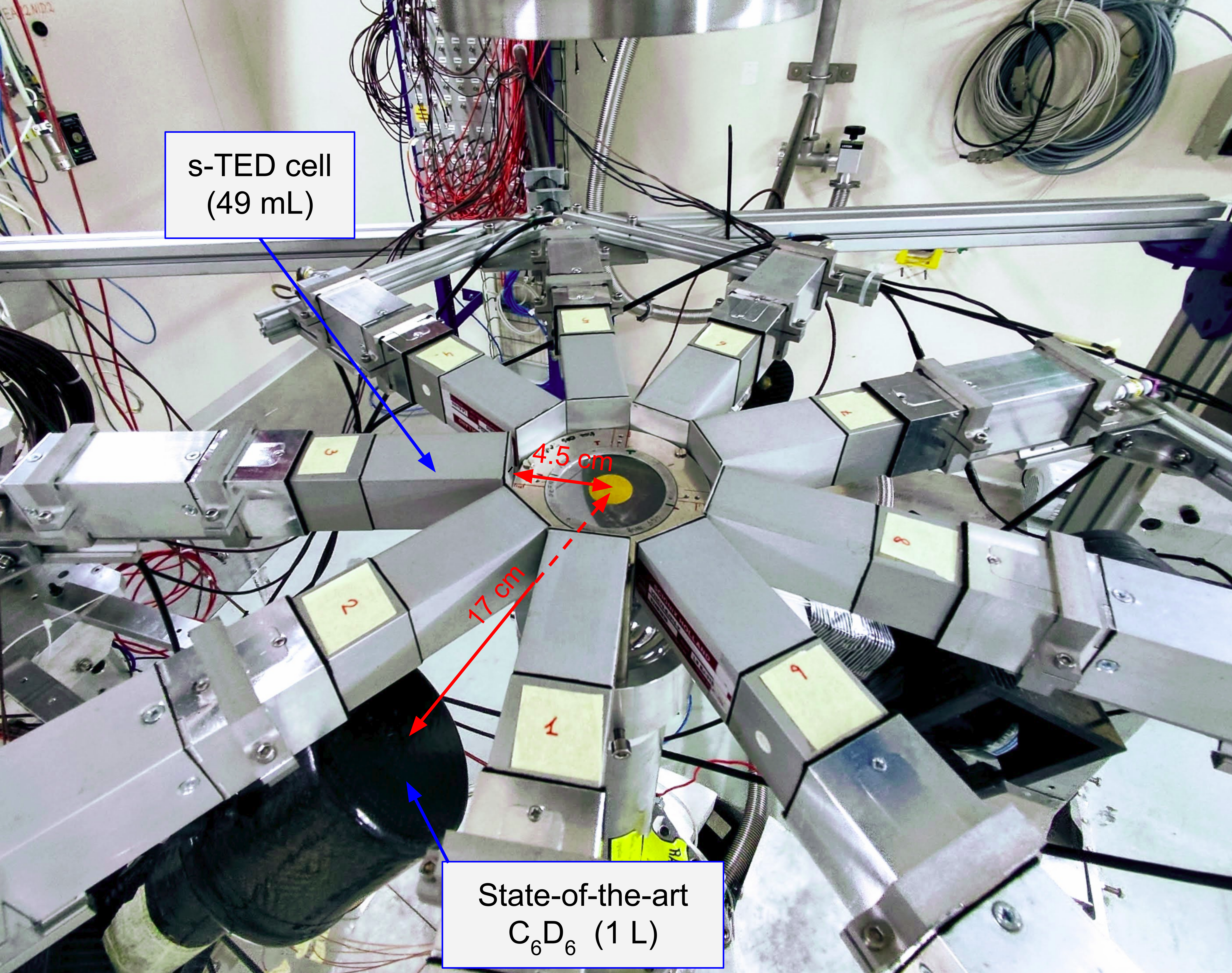}
  \includegraphics[width=1.0\columnwidth]{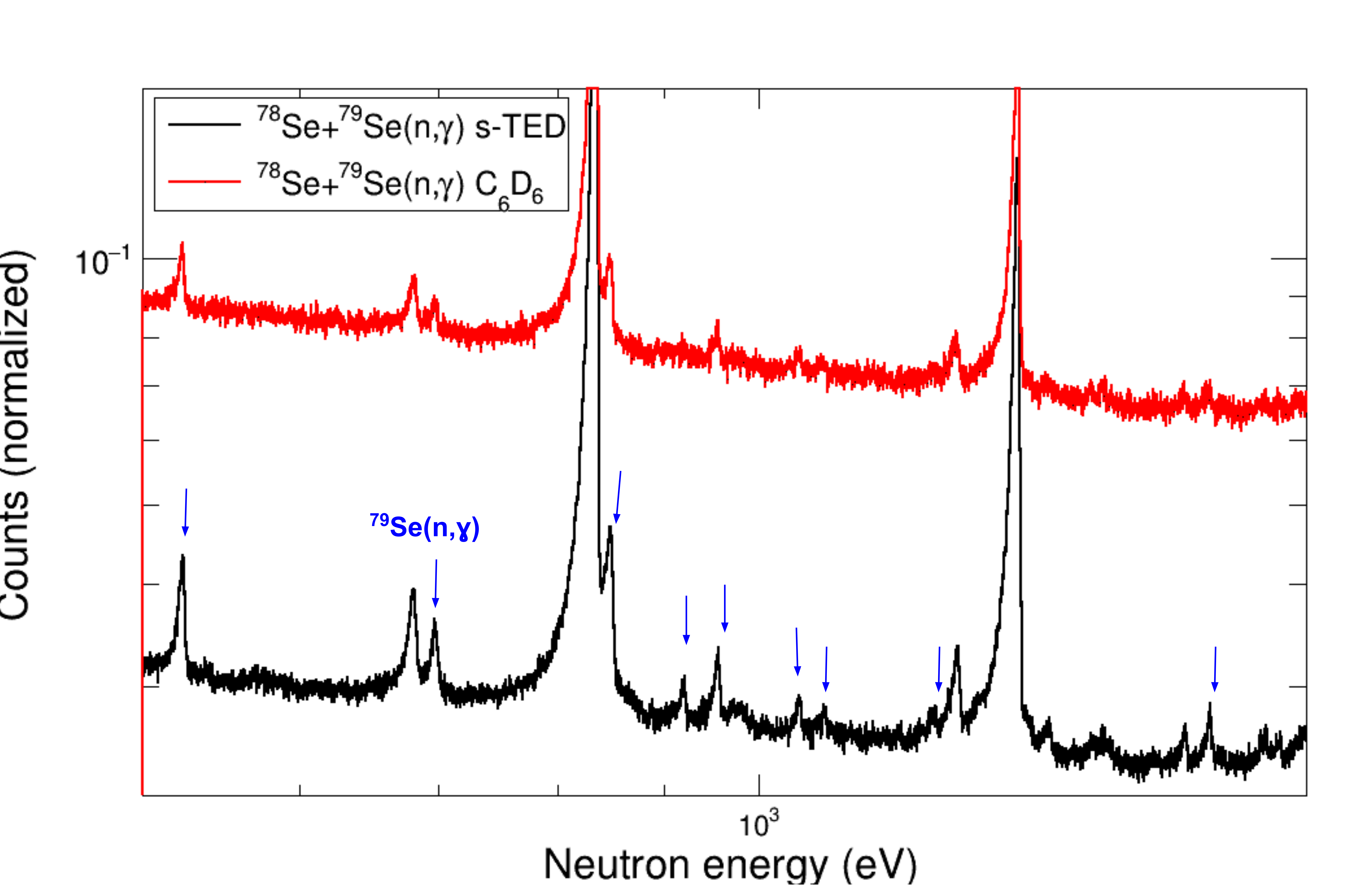}
  \caption{Top: Capture setup based on nine s-TED cells in ring-configuration and two conventional \cd~detectors used for the $^{79}$Se($n,\gamma$) experiment in 2022 at CERN n\_TOF EAR2. Bottom: Counting rates as a function of the neutron energy measured with the two setups normalized to the strongest resonance, highlighting some small  $^{79}$Se($n,\gamma$) resonances.}
  \label{fig:sTED}
\end{figure}

In order to overcome the aforementioned limitations of conventional \cd~detectors, a segmented array of small-volume \cd~detectors, so-called s-TED, has been developed~\cite{AlcayneND:22_b}. Each detection cell contains only 49~ml scintillation liquid, 12-20 times less than previous \cd~detectors. Nine of these cells were used in recent capture measurements at EAR2 on the unstable $^{94}$Nb~\cite{Balibrea21_b} and $^{79}$Se~\cite{Lerendegui21} in a compact-ring configuration around the capture sample shown in Fig.~\ref{fig:sTED}. This innovative setup minimizes the distance to the capture sample under study, and thus enhances the efficiency and SBR with respect to the larger conventional \cd~detectors~\cite{Balibrea22}. The improved SBR has been crucial to observe the weak $^{79}$Se(n,\g) resonances measured in a sample containing only 2.7~mg of $^{79}$Se embedded in 3.9~g of a eutectic lead-selenide ($^{208}$Pb$^{78}$Se) sample (see bottom panel of Fig.~\ref{fig:sTED}).

\section{Summary and outlook}\label{sec:summary}
The recent upgrade of the n\_TOF spallation target has allowed one to optimize its performance in terms of neutron flux and energy resolution in the two experimental areas. The first experimental results of (n,\g) measurements in the upgraded facility confirm the excellent energy resolution of EAR1 and show the clear improvement of the resolution in EAR2, which is key for both increasing the signal-to-background ratio and obtaining accurate cross sections.   

A large effort has been also done in detector R\&D at n\_TOF to solve existing limitations and further improve the capabilites for (n,\g) measurements. i-TED applies Compton imaging aimed at improving the signal-to-background ratio for measurements affected by large neutron-induced backgrounds. Exploiting the full potential of EAR2 required new detectors, such as s-TED, a new array of very small-volume C$_6$D$_6$ detectors, capable of dealing with the counting rate conditions and optimizing the signal-to-background ratio. These two novel detection systems have been already used in the challenging (n,\g) measurements on the unstable $^{94}$Nb~\cite{Balibrea21_b} and $^{79}$Se~\cite{Lerendegui21}.

After these upgrades, EAR1 is reinforced as an excellent facility for high resolution measurements using \cd~detectors on stable isotopes and samples with sufficient masses (>100~mg), being especially well suited for the determination of accurate resonance parameters over a broad energy range from 0.1~eV-300~keV, and measuring cross sections up to neutron energies of 1~MeV. The i-TED detector opens also the door to measure small (n,\g) cross sections on nuclei where (n,n) dominates. A promising future opens in EAR2 after the latest upgrades in energy resolution and detector performance. The 20-m vertical beam-line should consolidate as a reference facility on low mass (>100~$\mu$g) samples of unstable isotopes such as minor actinides or $s$-process branching points.

\section*{Acknowledgements}
This work has been carried out in the framework of a project funded by the European Research Council (ERC) under the European Union\'s Horizon 2020 research and innovation programme (ERC Consolidator Grant project HYMNS, with grant agreement No.~681740). This work was supported by grant FJC2020-044688-I funded by MCIN/AEI/ 10.13039/501100011033 and by European Union NextGenerationEU/PRTR. The authors acknowledge support from the Spanish Ministerio de Ciencia e Innovaci\'on under grants PID2019-104714GB-C21, FPA2017-83946-C2-1-P, FIS2015-71688-ERC, CSIC for funding PIE-201750I26. 


\end{document}